\documentclass[aps,prb,twocolumn,floats,showpacs]{revtex4}
\usepackage{epsfig}
\usepackage{graphicx}
\usepackage{hyperref}
\usepackage{amsmath}
\usepackage{amssymb}

\def\be{\begin{equation}}
\def\ee{\end{equation}}
\def\bea{\begin{eqnarray}}
\def\eea{\end{eqnarray}}

\def\ie{{\it i.e.\,}}
\def\etal{{\it et al.   }}

\def\<{\langle}
\def\>{\rangle}

\newcommand{\sguide}{\lrcorner \hskip-1.6pt\ulcorner}

\newcommand{\hide}[1]{}

\newcommand{\im}{\mbox{Im}}
\newcommand{\re}{\mbox{Re}}

\begin{document}

\title{Photonic structures with disorder immunity}

\author{E. Sadurn\'i}
\author{J. A. M\'endez-Berm\'udez}
\affiliation{Instituto de F\'isica, Benem\'erita Universidad Aut\'onoma de Puebla,
Apartado Postal J-48, 72570 Puebla, M\'exico}

\begin{abstract}

Periodic and disordered media are known to possess different transport properties, either classically or quantum-mechanically. This has been exhibited by effects such as Anderson localization in systems with disorder and the existence of photonic bandgaps in the periodic case. In this paper we analyze the transport properties of disordered waveguides with corners at very low frequencies, finding that the spectrum, conductance and wavefunctions are immune to disorder. Our waveguides are constructed by means of randomly oriented straight segments and connected by corners at right angles. Taking advantage of a trapping effect that manifests in the corner of a bent waveguide, we can show that a tight-binding approximation describes the system reasonably well for any degree of disorder. This provides a wide set of non-periodic geometries that preserve all the interesting transport properties of periodic media. 

\end{abstract}



\maketitle


\section{Introduction}

The study of transport properties in materials has important applications in the technological realm. With the advent of metamaterials [\onlinecite{veselago}], it has been demonstrated that such transport properties can be controlled by modifying the structure of solids. Both compositional and geometrical parameters play a significant role in the design of new materials. From the standpoint of physical phenomena, we distinguish the remarkable properties of photonic [\onlinecite{7,10}] and phononic crystals [\onlinecite{acoustic}], which emulate many features of electronic transport in solids: The presence of bandgaps, the appearance of conical points in the frequency spectrum of multilayered structures and the realization of effective Dirac equations in lattices [\onlinecite{sadurni2}] are just a few examples. 

The aforementioned properties usually depend on the periodicity of the assembled structures, including quantum-mechanical realizations in one dimension. Such examples of tight-binding chains can be found in polymers [\onlinecite{9}] and even in the modern construction of optical lattices [\onlinecite{inguscio}]. In this paper we go further and introduce disorder as another ingredient, with the purpose of constructing more flexible configurations. We show that the lowest energy band of waveguides with randomly oriented segments possesses the spectral properties of periodic structures, such as bandgaps, conical points (or Dirac points) and a conductance band. These are clear indications of robustness under disorder.

In connection with two-dimensional open systems  [\onlinecite{open}] and non-integrable billiards, we should mention that in some cases, randomly disposed boundaries and potentials [\onlinecite{mello}] give rise to wave-like manifestations of chaos [\onlinecite{stoeckmann}] in the form of level statistics [\onlinecite{mehta}] among other signatures. Additionally, Anderson localization [\onlinecite{anderson}] stands as one of the unmistakable signatures of disorder, affecting the corresponding transport properties. 

In our study we establish a result in the opposite direction, namely that low energy waves in systems with corners are {\it immune\ }to the effects of disorder, with no localized modes in the lowest frequency band and a non-vanishing conductance band below threshold. In fact, the examples studied in this paper recover one of the paradigms of integrability and solvability: The homogeneous tight-binding chain with nearest-neighbor interactions. 

The presence of corners in our disordered waveguides is of utmost importance. Chains of connected resonators can be proposed in many ways [\onlinecite{resonators1,resonators2}], but the shape of such resonators and the number of supported resonances inside them has an important effect in the complete system, perhaps in a rather uncontrolled manner. The fact that a corner connecting two leads at a right angle allows only one bound state can be used to our favor. The trapping mechanism of a corner is of a purely wave-like nature and manifests itself at the lowest part of the spectrum, where the wavelenghts are larger than the width of the guides. The presence of bound states in corners was first noted in [\onlinecite{schult}] and their description was developed in [\onlinecite{sadurni}] by means of conformal maps and effective potentials.


In this paper we study the spectrum, eigenfunctions and dimensionless conductance of disordered waveguides with corners. As an important result we obtain transmission below the threshold of a straight waveguide (section III), forming a low frequency band located around the single trapped mode of an isolated corner or L-shaped waveguide and producing a gap which extends from the edge of the lowest band to the straight-guide propagation threshold. Several realizations of two dimensional pipes are obtained by varying the orientations of their segments, showing thus the robustness and flexibility of the system. Then we give an analytical explanation of these effects by finding the effective interaction of a wave with a corner through a conformal map and proceed to connect corners in tigh-binding schemes (section IV). We finish with a summary and an outlook (section V).

\section{Definition of our system}

We are interested in the transport properties of two-dimensional waveguides with corners bent at right angles and randomly oriented segments. There are two models which can be considered. See Figs.~\ref{examples} and \ref{examples2}. We describe their geometry as follows.

\subsubsection{Model 1}
This model is built by assembling blocks with the forms $\sqcup$ and $\sqcap$. These blocks are randomly alternated as one moves along the array, with the only condition the array contains no loops. The resulting configurations are almost horizontal, minimizing the space. We relax this condition in Model 2. The parameters are the following: $d$ is width of the waveguide, $L$ is the length of the straight segments in units of $d$, $N$ is the number of unit cells ($\sqcap$ or $\sqcup$) forming the guide, the quantity $p$ is the probability of finding $\sqcup$ cells (for many realizations this is roughly the ratio of $\sqcup$ to $\sqcap$ cells).

\begin{figure}[t]
\centerline{\includegraphics[width=8cm]{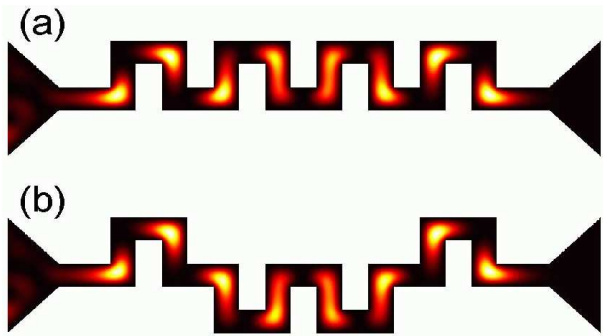}}
\caption{ \label{examples}(Color online) Wave amplitudes $|\Psi(x,y)|^2$ for influx coming from the left lead 
in waveguides (Model 1) with $L=3d$, $N=4$, $D=5d$. In (a) we set $p=0$, while in (b) $p=0.5$. In both cases the resonant energy $E=0.9412 E_t$ was used.}
\end{figure}

\begin{figure}[t]
\centerline{\includegraphics[width=8cm]{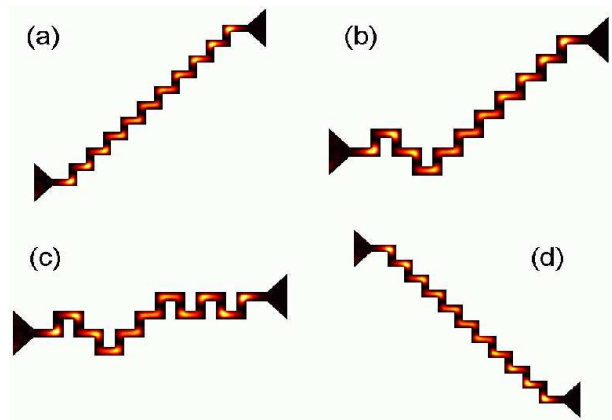} }
\caption{\label{examples2}(Color online) Wave amplitudes $|\Psi(x,y)|^2$ for influx coming from the left lead 
in waveguides (Model 2) with $L=3d$, $N=10$, $D=5d$. (a) $p=0$, (b) $p=0.2$, (c) $p=0.4$, (d) $p=1$. 
In all cases the resonant energy $E=0.9504 E_t$ was used.}
\end{figure}

Examples are shown in Fig.~\ref{examples}. The waveguide is
connected to two semi-infinite collinear leads of width $D$. The leads are 
attached to the waveguide by means of triangular contacts, avoiding strong diffractive effects. The width of the leads can be chosen arbitrarily, as long as waves of a wavelength larger than $d$ can be supported. It suffices to take $D=5d$.

\subsubsection{Model 2}

We consider waveguides with the same parameters as for Model 1, however
here the cells forming the waveguide have forms L or $\Gamma$, allowing more flexible configurations. Now 
$p$ is the probability of finding cells of type L. As before, we fix $D=5d$.
See waveguide examples in Fig.~\ref{examples2}.

\subsection{Boundary value problem}

We focus on the solutions of the stationary Schr\"odinger wave equation subjected to Dirichlet boundary conditions which are defined by the shape of our waveguides. Our results shall equally hold in settings involving electromagnetic waves [\onlinecite{stoeckmann}] or acoustic waves in the linear regime [\onlinecite{elasticity}], as we are dealing with the Helmholtz equation in a hollow guide. For a quantum particle of mass $\mathbf{m}$ we have

\bea
\left(\frac{\hbar^2}{2 \mathbf{m}}  \nabla_{x,y}^2 + E \right) \psi(x,y) = 0
\label{1.1}
\eea
and $\psi(x,y)=0$ at the boundary.
In principle, we can infer the properties of our system by sending a wave of fixed energy through one of the leads (of width $D$) and finding the scattering amplitudes. 
The energy is given by
\begin{equation}
E = \frac{\hbar^2}{2\mathbf{m}} \left( k_m^2 + \frac{m^2\pi^2}{D^2} \right),
\end{equation}
where $k_m$ and $m\pi/D$ are, respectively, the
longitudinal and transversal momentum components of the total wave vector with
magnitude $K=\sqrt{2{\mbox{\bf m}}E}/\hbar$. Our purpose is to explore the lowest energy region of the system. Therefore it is convenient to express the energy in 
the units $E/E_t$, which normalize our quantities with respect to the energy of the lowest mode of the guide, \ie the threshold energy $E_t=(\hbar^2/2\mathbf{m})(\pi^2/d^2)$.

\section{Numerical results}

Using finite element methods we compute the scattering matrix ($S$-matrix)
which has the form
\begin{equation}\label{notacion}
S = \left(\begin{array}{cc}
      t & r' \\
      r & t' 
    \end{array} \right) \ . 
\end{equation}
The symbols $t$, $t'$, $r$, and $r'$ are $M\times M$ transmission and reflection 
matrices, where $M$ is the highest mode given by the largest $m$ beyond which 
the longitudinal wave vector
\[ 
k_m=\sqrt{K^2-m^2\pi^2/D^2}
\]
becomes complex.
Once the $S$-matrix is known we calculate the dimensionless 
conductance (see [\onlinecite{landauer,imri,bu}] and its application to mesoscopic systems in [\onlinecite{mello}]).
\begin{equation}
T = \mbox{Tr}(tt^{\dagger}) \ .
\end{equation}

\subsection{Results for Model 1}

\begin{figure}[b]
 \includegraphics[width=8cm]{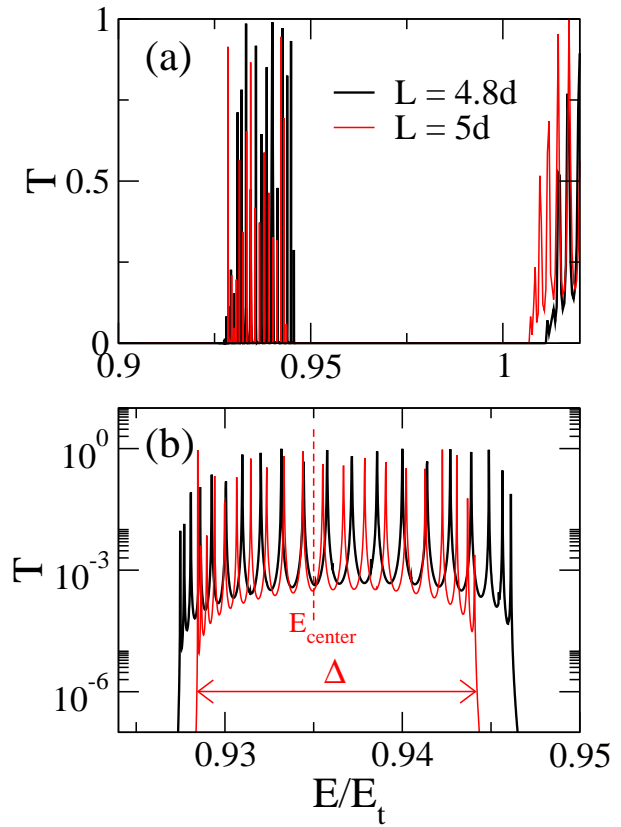}
\caption{ \label{G1}(Color online) (a) Conductance $T$ as a function of $E$ for waveguides (Model 1) with 
$N=10$, $p=0.5$, and $L=4.8d, 5d$. A band below threshold appears for both values of the intersite distance. (b) Amplification of (a) in the lowest conductance band. The bandwidth $\Delta$ increases as the corners approach each other, while the center $E_{\rm{center}}$ hardly moves when the geometry is altered.}
\end{figure}

In Fig.~\ref{G1} we show the conductance $T$ as a function of $E$ for waveguides 
with $N=10$, $p=0.5$, and $D=5d$. For comparison purposes we fix a single random 
sequence of unit cells and present results for $L=4.8d$ and $L=5d$.
We note that the conductance plots change importantly for different values of $L$, but all the arrays 
display local maxima (resonant energies) in the region below threshold. The existence of a gap ranging from $E \sim 0.95 E_t$ to $E \sim E_t$ is evident in Fig.~\ref{G1}(a).

In Fig.~\ref{G1}(b) we have amplified the region of the conductance band. Here we observe an increase of the bandwidth as the geometric parameter $L$ decreases, a behavior that can be interpreted as an increase of the coupling between corners in a tight-binding regime. This shall be explained in further sections. The number of resonant peaks coincides with the number of corners in the array.

\begin{figure*}[t]
\centerline{\includegraphics[width=13cm]{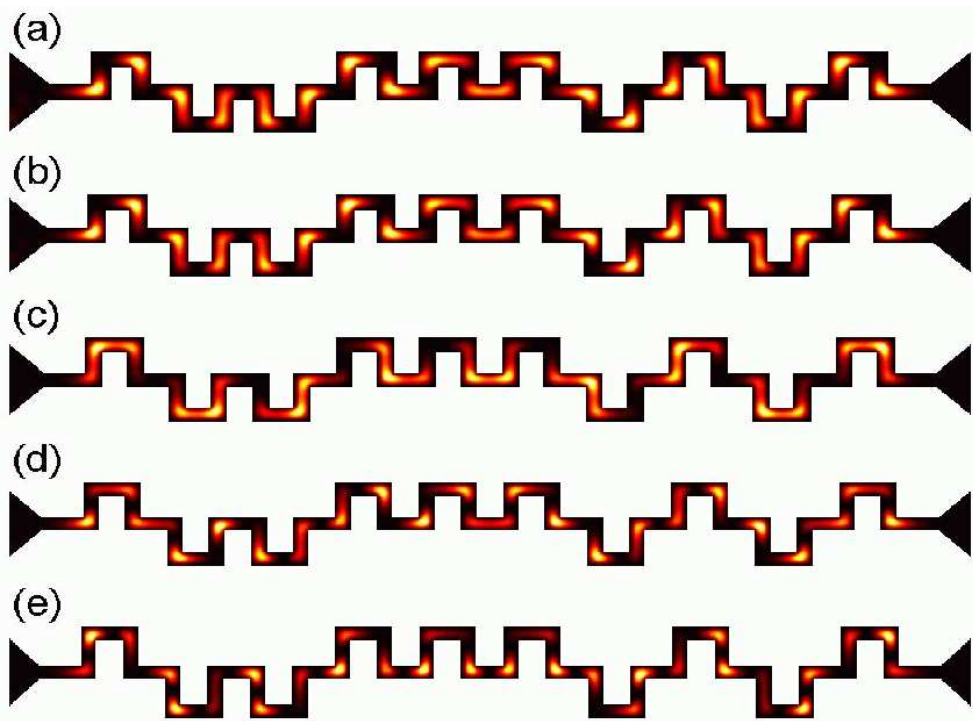}}
\caption{ \label{wfs1}(Color online) Resonant wave amplitudes $|\Psi(x,y)|^2$ for influx coming from the left lead 
in waveguides (Model 1) with $N=10$, $D=5d$, and $p=0.5$. From top to bottom $(L,E/E_t)=(3d, 0.9412)$,
$(3.2d, 0.9384)$, $(3.4d, 0.9184)$, $(3.6d, 0.944)$, and $(3.8d, 0.9552)$. See Fig.~\ref{G1}.}
\end{figure*}

In Fig.~\ref{wfs1} we plot wave intensities $|\Psi(x,y)|^2$ at resonant 
energies for waveguides similar to the ones used in Fig.~\ref{G1}. There is a visible accumulation of wave intensities near the corners, becoming more pronounced as $L$ increases. The accumulation, however, does not occur in every corner: The intensities display the behavior of Bloch waves from corner to corner, despite the fact that the arrays are non-periodic, showing a quasi 1d propagation in a periodic medium.  

We have found that the effect of disorder in the conductance band under scrutiny is minimal, as there are small differences in $T$ for different configurations of the guides parameterized by the value of $p$. See section III.D.

\subsection{Results for Model 2}

As it is evdent form  Fig.~\ref{examples2}, the resulting configurations are more flexible than those of Model 1 (as long as the array does not intersect itself). This model allows to connect endpoints at arbitrary heights. The results for the conductance $T$ as a function of $E$ for waveguides 
with $N=10$ is quite similar to those previously described for Model 1. The resulting conduction band has resonant peaks that numerically approach $T=1$ and formally reach $T=1$. The number of peaks of $T$ in the band for these configurations coincides with $2N$.

As before, we observe that once $N$ and $L$ are fixed the conductance plots are almost 
the same for different values of $p$. A more detailed description is given in section III.D.
Now, the conductance band can also be analyzed as a function of $L$, increasing from $2.5d$ to $5d$.
A detailed analysis shows that (i) the bands are asymmetric for small $L$ and 
the bandwidth $\Delta$ decreases with increasing $L$, see Fig.~\ref{G7}(a); and 
(ii) the band center $E_{\rm{center}}$ moves to smaller values of $E$ with increasing $L$, see 
Fig.~\ref{G7}(b). See section III.C.

From our calculations we can also conclude that the differences in the conductance profiles 
between ordered and disordered waveguides (i) are larger the smaller the value of $L$
is; and (ii) are larger on the left side of the bands.

\subsection{Structure of the lowest conductance band}

\begin{figure}[h!]
\centerline{ \includegraphics[width=8.5cm]{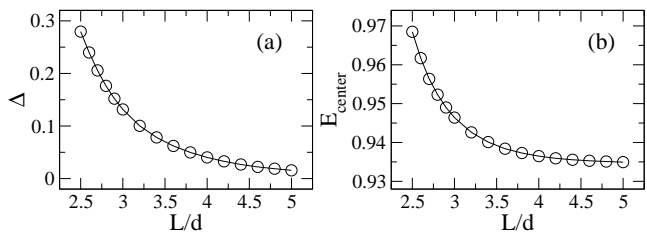}}
\caption{ \label{G7} (a) Bandwidth $\Delta$ and (b) band center $E_{\rm{center}}$ as functions of $L/d$, for 
waveguides of Model 2 with $N=10$, $p=0$, and $D=5d$. Both quantities decay exponentially with the distance. While the bandwidth $\Delta$ ranges from $0$ to $0.3 E_t$, the center varies only in the range $0.93 E_t  <E_{\rm{center}}< 0.97 E_t $.}
\end{figure}

Our goal is to study the low frequency bands formed by the conductance below threshold. To this end we analyze numerically the variation of the bandwidth and the position of the center of the band as functions of the geometry. Scale transformations of the guide in the form $d \rightarrow \lambda d$ and $L \rightarrow \lambda L$ modify the spectrum trivially by an overall scale of $\lambda^{-2}$. In order to modify the structure of the resulting spectrum we vary only one geometrical parameter, in our case $L$. In this way we find the behavior of the band when the distance between corners increases. We take $\Delta$ as the bandwidth defined by the difference of the two energies below $E_t$ at which the quantity $T<10^{-6}$. The center of the band $E_{\rm{center}}$ is simply taken as the energy which divides the number of peaks in two equal parts. See Fig.~\ref{G1}.

In Fig.~\ref{G7} we find an exponential decrease of the bandwidth with the distance between corners. The values range from $0$ (long distance interaction) to $0.3 E_t$ (strongest interaction at $d=2.5$). This supports the idea that two coupled corners should be suffient to describe the coupling between sites and its behavior as a function of the distance between them. On the other hand, the center of the band shows little variations: for long distances we have a center at $0.93 E_t$, while for $d=2.5$ the center approaches the threshold at $ 0.97 E_t$. It should be noted that at long distances, all levels tend to be degenerate at $0.93 E_t$. The meaning of this energy shall be ellucidated in further sections, where we show that a single corner at a right angle can support only one bound state lying at such an energy. We shall use these features in order to establish the validity of a nearest-neighbor interaction in the theoretical description.

\subsection{Small effects at the edge of the band}

\begin{figure}[t]
\centerline{ \includegraphics[width=8cm]{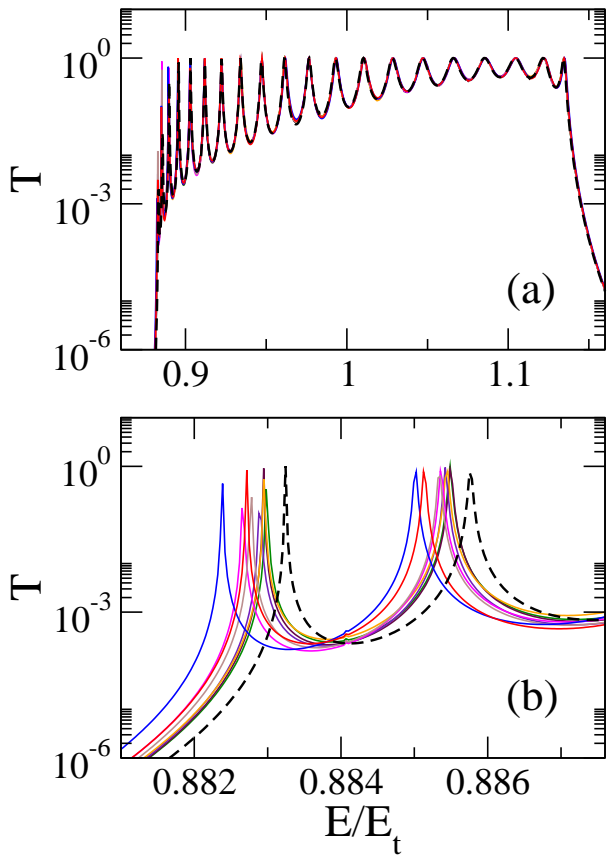}}
\caption{ \label{G8}(Color online) (a) Conductance $T$ as a function of $E$ for waveguides 
(Model 2) with $N=10$, $L=2.5d$, and $D=5d$ for $p=0$ (dashed curve) and 
8 waveguide realizations with $p=0.1,...,0.5$ (color curves). (b) 
Deviation of resonant peaks of the conductance as a function of the number of defects. An increasing disorder produces a small deviation of the peaks at the edge of the conductance band. The largest shift is of the order $10^{-3} E_t$ for $p=0.5$ corresponding to the blue curve.}
\end{figure}

We have seen that the effects produced by disorder are small. Nevertheless such effects can be distinguished by a close inspection of the conductance at the edges of the bands. Here we analyze numerically the consequences of introducing disorder in the arrays. In Fig.~\ref{G8}(a) we show the peaks forming the conductance band, with curves of different color for values of $p=0.1,...,0.5$. The effects due to disorder are not visible at this energy scale, even at a separation distance of $L=2.5 d$ between corners. In Fig.~\ref{G8}(b), the scale has been increased, showing the effects on the position of the peaks at the lower edge of the conductance band. The shift of the peaks occurs downwards and it increases with the value of $p$, widening the conductance band by amounts less than $10^{-3} E_t$.

\section{Theoretical description}

Our aim is to describe the effects numerically found in the previous analysis. For this purpose, we follow three steps. 1) We describe the trapping mechanism of a single corner, ensuring the existence of one bound state below threshold. This shall be done by means of a conformal map, giving rise to an effective interaction in the form of a position-dependent mass (in case of a quantum wire) or an effective dielectric function (for electromagnetic waves). 2) We analyze the interaction between two corners coupled in two different configurations, namely $\sguide$ and $\sqcap$. As the spectrum of these two systems are approximately equal (even when distances between corners are reduced), we conclude that the differences can be treated perturbatively. 3) We introduce a nearest-neighbor tight-binding model where the atomic sites are represented by corners. The effect of disorder is described perturbativley, explaining the effects at the edge of the conductance band.

\subsection{The trapping mechanism of a corner}

\begin{figure}[t]
\begin{tabular}{cc}
 \includegraphics[width=5cm]{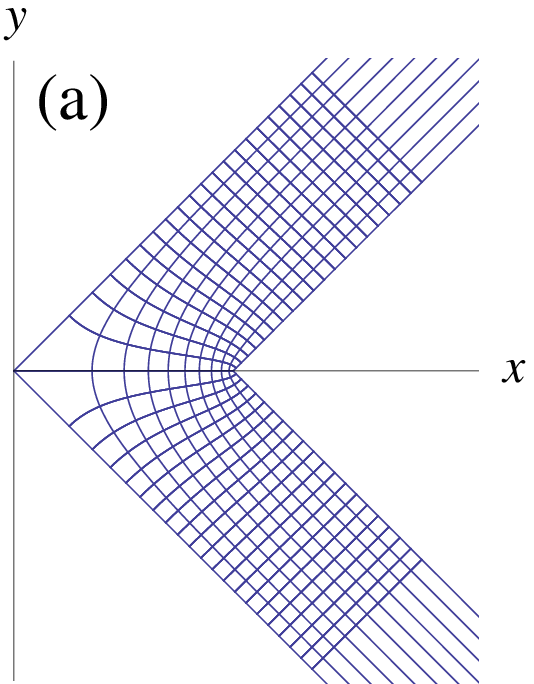}
& \includegraphics[width=3cm]{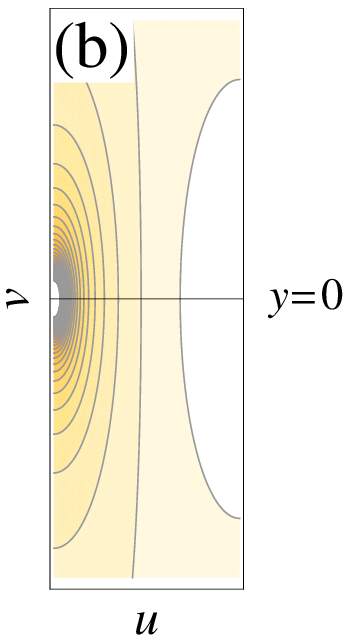}
\end{tabular}
\caption{ \label{fig0}(Color online) (a) Conformal coordinates obtained from (\ref{eq3}) in the form $x(u,v)+iy(u,v)= \csc (3 \pi/8) I \left(\sin^2 \left[(u+iv)\pi/2\right],1/4,3/4 \right)$. Its behavior near corners shows an abrupt change of the contour density, related to the Jacobian of the transformation. (b) Contour plot of the Jacobian as a position-dependent mass, its value being minimal around the position of the internal corner and approximately unity along the arms.}
\end{figure}

The L-shaped waveguide can be transformed into a straight one by means of a conformal map. The resulting Helmholtz equation acquires a position-dependent factor in the Laplace operator. For quantum-mechanical waves this can be interpreted as a position-dependent mass, whereas for components of electromagnetic fields this can be thought as an effective dielectric medium. We proceed as follows. 
We solve the stationary Schr\"odinger (or Helmholtz) equation as a Dirichlet boundary value problem defined by our waveguide in Fig.~\ref{fig0}. We use units $1=\hbar^2/2\mathbf{m}$ and coordinates $x,y$ to write

\bea
\left[ \nabla_{x,y}^2 + k^2 \right] \phi(x,y) = 0, \qquad \phi|_{\partial \Omega} = 0,
\label{eq1}
\eea
where $\Omega$ is the interior of the array. In the following we describe our conformal map in order to find a position-dependent mass or an effective dielectric function of the coordinates.  The obvious choice for a conformal set of coordinates is a function which maps an infinite straight strip into a bent waveguide. Let $F$ be an analytic function such that $u= \re[F(x+iy)], v=\im[F(x+iy)]$. This leads to a transformation 

\bea
\nabla_{x,y}^2 = \frac{\partial(u,v)}{\partial(x,y)} \nabla_{u,v}^2, 
\eea
where the Jacobian appears as a prefactor and satisfies 

\bea
\frac{\partial(u,v)}{\partial(x,y)}=\left. \left| \frac{dF(z)}{dz} \right| ^2  \right|_{(u,v)}, 
\eea
where $z=x+iy$. The boundary value problem (\ref{eq1}) for the function $\psi(u,v) \equiv \phi(x(u,v),y(u,v))$ has the equivalent forms

\bea
\left[ \nabla_{u,v}^2 + \eta^2(u,v) k^2  \right] \psi(u,v) = 0, \nonumber \\ \psi(0,v)=\psi(1,v)=0,
\label{eq2}
\eea
with an effective dielectric function $\eta(u,v)= \sqrt{\frac{\partial(x,y)}{\partial(u,v)}}$ or 
\bea
\left[  \frac{1}{\mu(u,v)}\nabla_{u,v}^2 + E \right] \psi(u,v) = 0, \nonumber \\ \psi(0,v)=\psi(1,v)=0,
\label{eq2.1}
\eea
with an effective mass  $\mu(u,v)= \frac{\partial(x,y)}{\partial(u,v)} $.

The passage from the wavefunction $\psi(u,v)$ to a normalizable $\phi(x,y)$ implies the use of the Jacobian mentioned above. Therefore, the old wavefunction $\phi(x,y)$ would also satisfy the boundary conditions if the square root of the Jacobian does not contain strong singularities according to $\lim_{u \to 1,0}  \psi(u,v)|\frac{dF}{dz}|_{(u,v)}=0 $. The explicit form of $F$ can be chosen in many ways. However its behavior near the corners is universal, since the angle formed by the walls of the array represents a branchcut of the map and it determines uniquely the rational power $q$ appearing in $F \sim (z-z_0)^q$, where $z_0$ is a vertex on the boundary. Here we find it convenient to construct $F$ by means of a composition of two Schwarz-Christoffel transformations (see Fig.~\ref{fig0}). One of them maps an infinite strip of unit width to the semiplane and the other maps the semiplane to a tilted trigon [\onlinecite{conformal}]. For waveguides bent in a right angle we have 

\bea
F(z) = u+iv = \frac{2}{\pi} \arcsin \left[ \sqrt{I^{-1}\left(z\sin \left(\frac{3\pi}{8}\right),\frac{1}{4},\frac{3}{4} \right)} \right] , \nonumber \\
\label{eq3}
\eea
where $I^{-1}$ is the inverse of the regularized Beta function [\onlinecite{gradshteyn}]. For further details and more general bending angles, we refer the reader to [\onlinecite{sadurni}].

The spectrum of this system can be obtained by solving (\ref{eq1}) or its equivalent forms (\ref{eq2}) and (\ref{eq2.1}). It is composed by a single bound state lying below the propagation threshold of the straight segments and a continuum of energies above such a threshold. Henceforth we shall refer to the bound state energy and threshold energy as $E_0$ and $E_t$ respectively. It has been shown numerically and analytically [\onlinecite{exnerl}, \onlinecite{schult}] that $E_0 \approx 0.925 E_t$ and that the wave function of the single bound state decays exponentially along the arms with a decay length $\lambda \sim \sqrt{E_0}$ [\onlinecite{sadurni}].

\subsection{Two connected corners: $\sguide$ vs $\sqcap$ pipes}

The results above suggest that each corner can act as an atomic orbital in a lattice, with the possibility of connecting many corners to form a wide class of structures. In order to ensure the rise of a tight-binding chain, we focus first on the interaction between two corners in two different configurations: $\sguide$ guides and $\sqcap$ guides. These two systems can be regarded as two-level atoms concerning their spectrum. They shall be studied numerically in order to show a level splitting (or coupling) which decays exponentially with the distance between corners. The degenerate levels approach the single bound state level of the L-guide, regardless of the orientation of the couplings ($\sguide$ or $\sqcap$). The small differences in the properties of $\sguide$ and $\sqcap$ shall then be exploited in more complex assemblies containing them as building blocks. In this way, the bandwidth and bandcenter of a disordered chain can be shown to be almost independent of the realization and can be further related to the interaction between two corners (we can refer to them indistinctly as atomic orbitals or lattice sites).

\begin{figure}[t]
\begin{tabular}{cc}
\includegraphics[width=4cm]{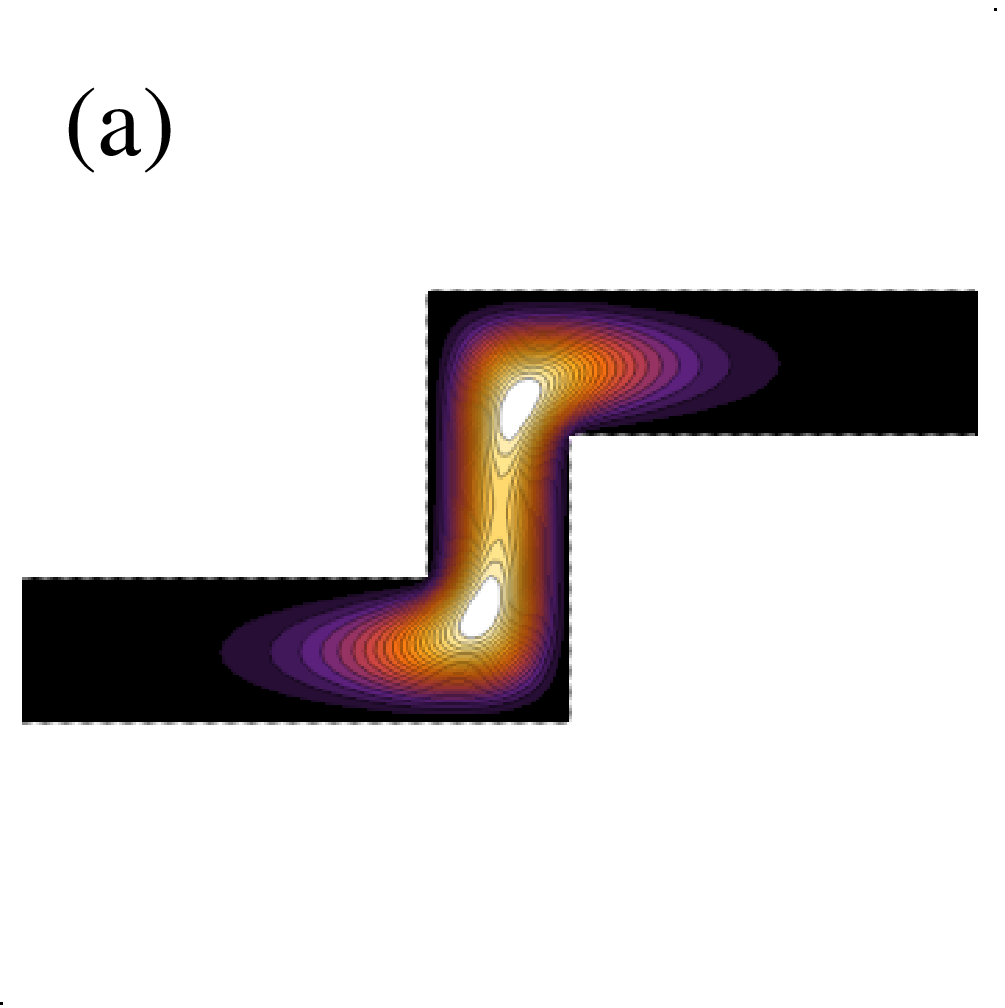} &
\includegraphics[width=4cm]{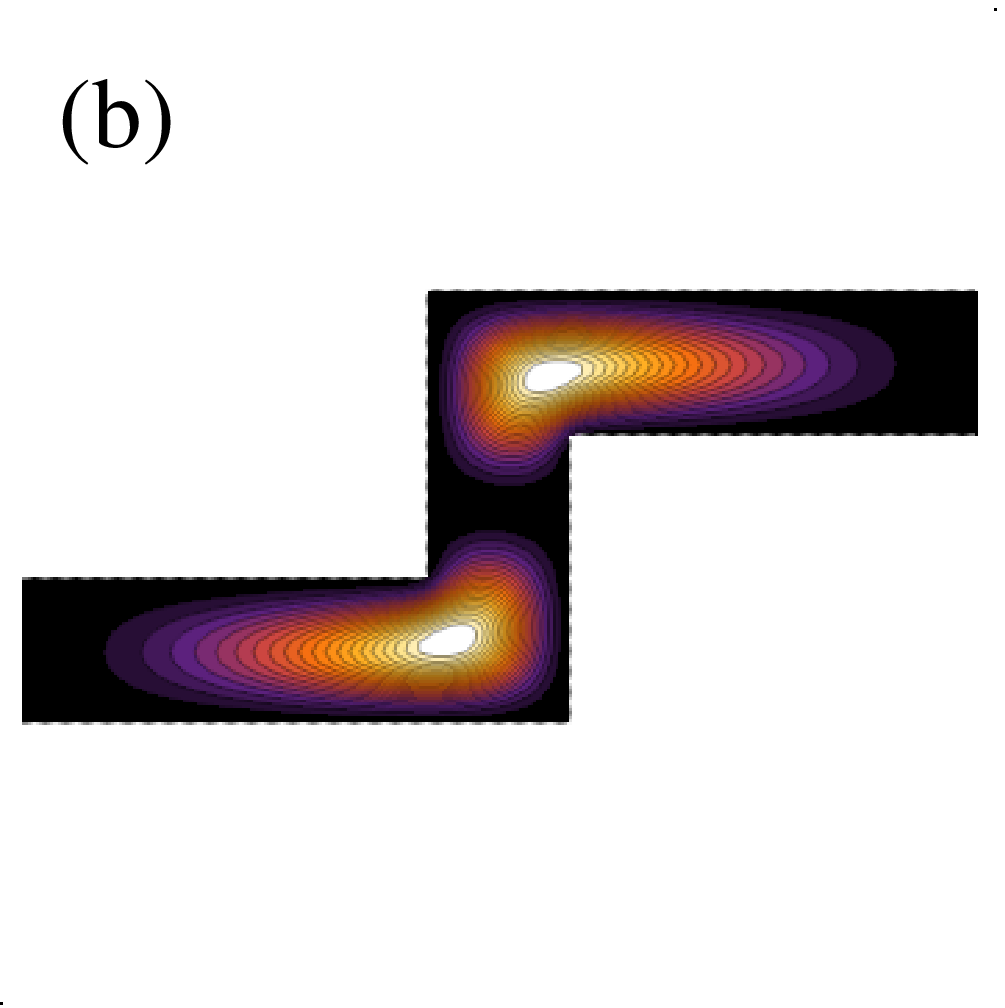} \\

\includegraphics[width=4cm]{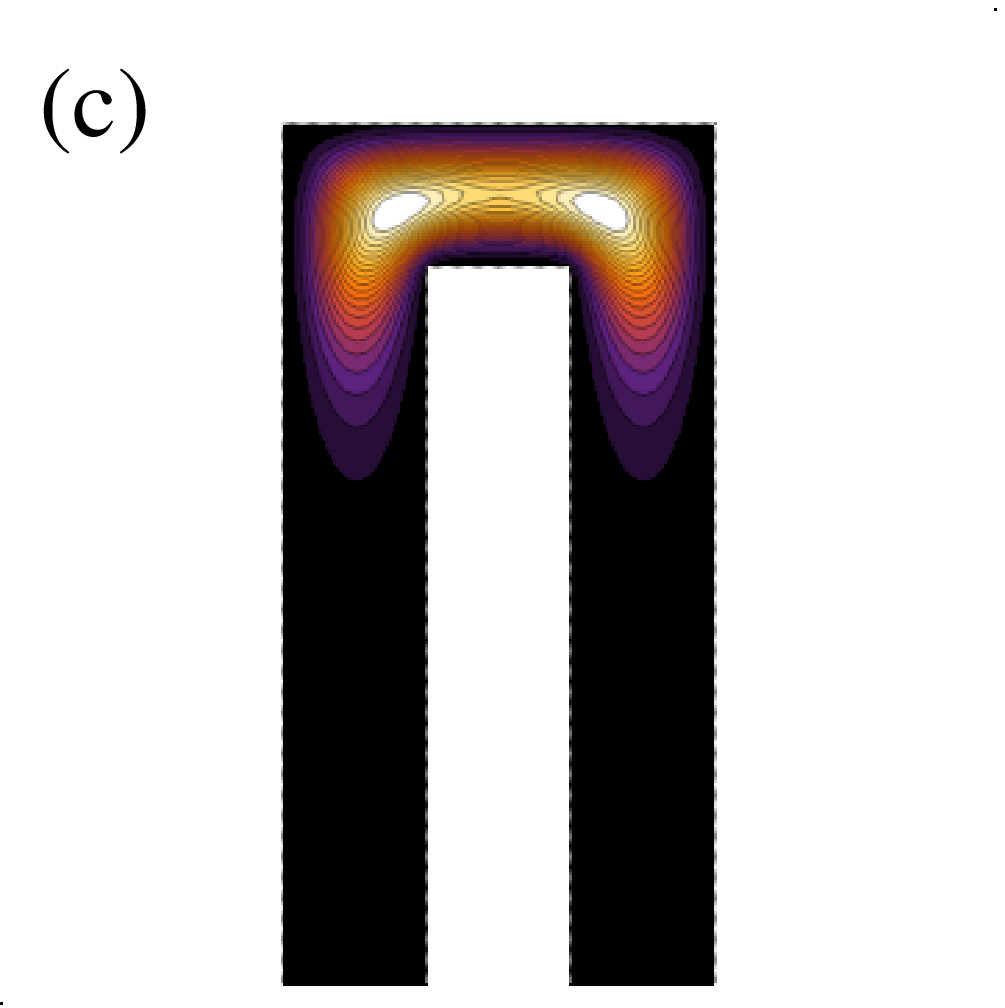} &
\includegraphics[width=4cm]{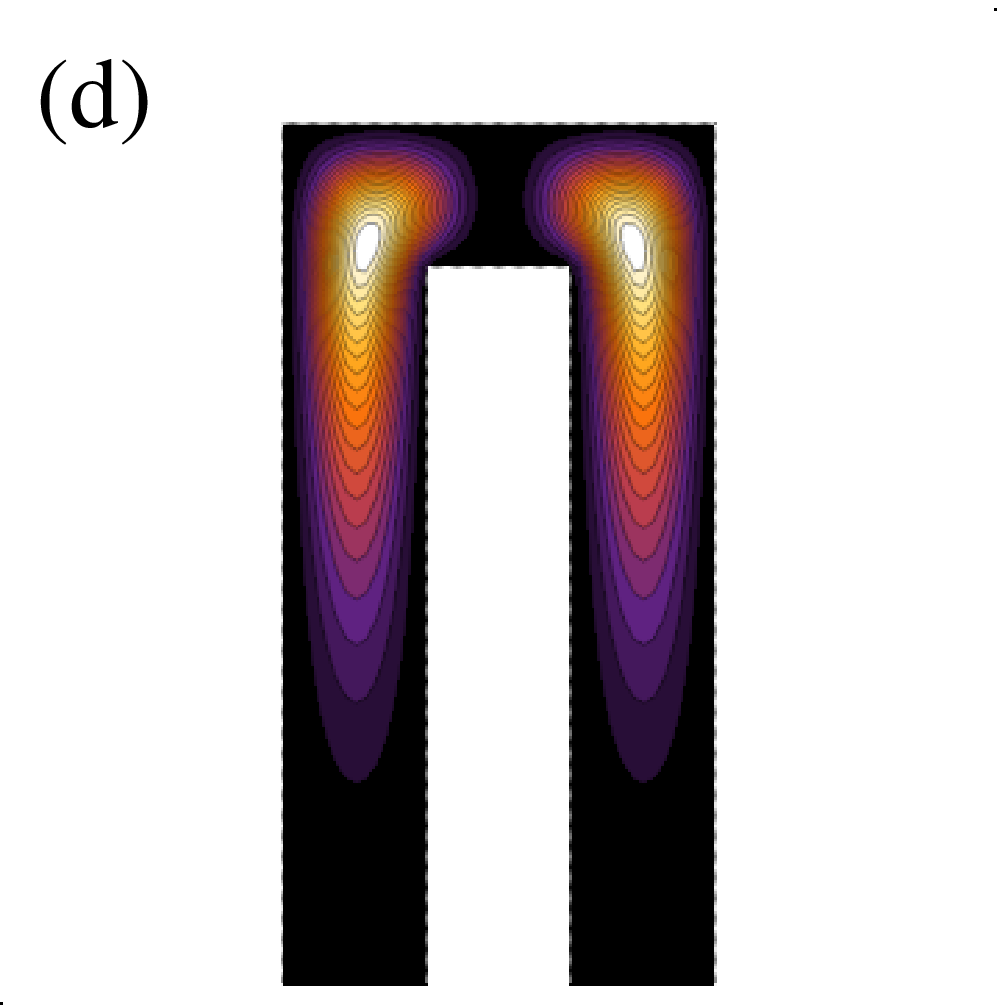}

\end{tabular}
\caption{ \label{SandU}(Color online) Bound states of two interacting corners with $L=3d$. (a) Symmetric state for $\sguide$ guide. (b) Antisymmetric state for $\sguide$ guide. (c)  Symmetric state for $\sqcap$ guide. (d) Antisymmetric state for $\sqcap$ guide.}
\end{figure}

\begin{figure}[t]
\centerline{\includegraphics[width=8cm]{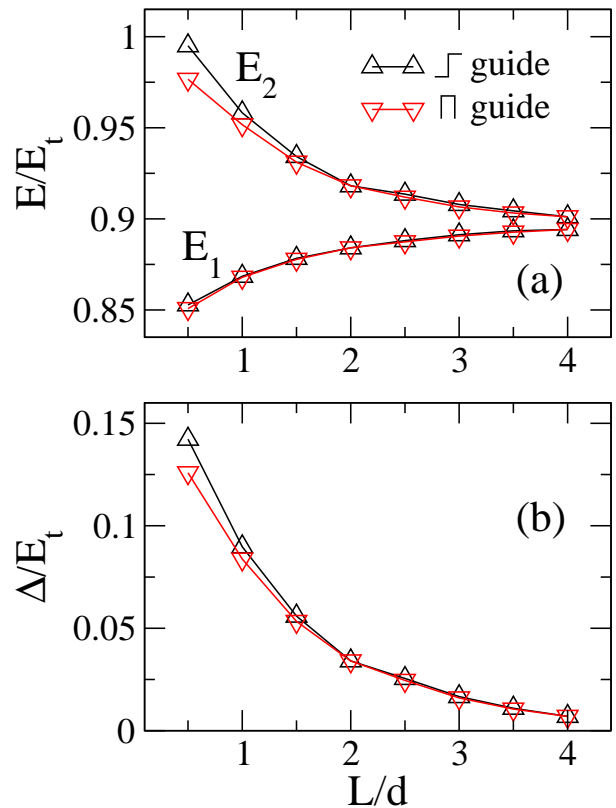}}
\caption{ \label{levels} (Color online) (a) Level collapse as a function of the distance $L$ between corners. The decay of the coupling $\Delta$ follows an exponential law. (b) Energy difference of the two low-energy levels for each configuration. Black curve: $\sguide$ guide, Red curve: $\sqcap$ guide. Discrepancies are almost negligible, even near $L/d=0.5$, where the difference between the splittings of the two systems is maximal \ie $\Delta \epsilon / 2 \sim 0.017 E_t$.}
\end{figure}

We solve the stationary Schr\"odinger (Helmholtz) equation in two dimensions with the geometries $\sguide$ and $\sqcap$ and Dirichlet boundary conditions. The finite element method finds all eigenfunctions and energies, particularly those lying at the lowest part of the spectrum. We show in Fig.~\ref{SandU} the resulting wave functions. The low-energy levels $E_1$ and $E_2$, and the level splitting (or effective coupling) denoted by $\Delta = 2(E_2-E_1)$, can be given as functions of the distance $L$ between corners. The factor of $2$ has been introduced in order to compare $\Delta$ with our previous definition of bandwidth. The results are shown in Fig.~\ref{levels}. The exponential decay of the level splitting with the distance $L$ ensures that the coupling of two L-waveguides emulates two-level atoms for both $\sguide$ and $\sqcap$ guides. The small differences between the energies for $\sguide$ and $\sqcap$ configurations guarantee the immunity to disorder of structures that are randomly built from these blocks. 
Given our numerical results, we can establish a model hamiltonian for energies exclusively below threshold. The two-level $H$ has the form

\bea
H_{\sguide} = \left(\begin{array}{cc} E_0 & \frac{\Delta}{4} \\ \frac{\Delta}{4}  & E_0  \end{array} \right)
\label{modelS}
\eea
for the $\sguide$ configuration and
\bea
H_{\sqcap} = \left(\begin{array}{cc} E_0 & \frac{\Delta}{4}(1 + \epsilon) \\ \frac{\Delta}{4}(1 + \epsilon) & E_0  \end{array} \right)
\label{modelU}
\eea
for the $\sqcap$ shape. The small differences in the level splittings of the two configurations are given by $\Delta \epsilon / 2$. The hamiltonian (\ref{modelU}) can be cast as a perturbation of (\ref{modelS}) \ie  $H_{\sqcap} = H_{\sguide} + \epsilon V $ with $V$ an off-diagonal potential. We shall use this potential in the construction of chains with blocks of the $\sguide$ and $\sqcap$ types.

\subsection{The tight-binding chain with disorder}

\begin{figure}[t]
\begin{tabular}{c}
\includegraphics[width=8cm]{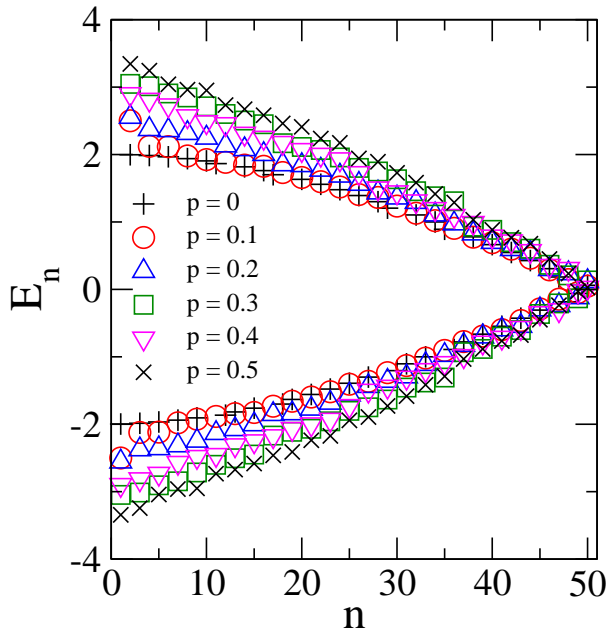}
\end{tabular}
\caption{ \label{random}(Color online) Numerical energies as a function of the eigenvalue number $n$ for a chain of $50$ sites randomly coupled according to (\ref{hamiltonian2}). The parameters are $\Delta=4$ and $\epsilon = 0.1$. The colors from red to blue indicate realizations corresponding to values of $p$ increasing in steps of $0.1$. The behavior of our disordered chain is consistent with numerical calculations of conductance peaks, as the bandwidth increases monotonically with the number of perturbations or defects, modifying the edge of the band. On the other hand, the conical point of the spectrum at eigenvalue $50$ is immune to off-diagonal disorder due to the swapping symmetry. See the text below (\ref{shiftp}).}
\end{figure}

Now that we have shown that the differences between the coupling of corners in $\sguide$ and $\sqcap$ shapes are small, we can construct a disordered tight-binding chain with many of these blocks. In the following we shall use states localized at the corners as a basis for the Hilbert space of the wave operator for energies below threshold. Denoting the $n$-th corner state by $| n \>$ and its localized wavefunction by $\xi(x-x_n)=\<x|n\>$, we propose the lowest energy band hamiltonian of a homogeneous chain of $N$ corners as

\bea
H= E_0 + \frac{\Delta}{4} \sum_{n=1}^{N-1} \{|n\>\<n+1| + \rm{h.c.} \},
\label{hamiltonian1}
\eea
where only blocks of the $\sguide$ type (comprising two corners) appear. If $N$ is sufficiently large, we obtain Bloch waves as eigenvectors of (\ref{hamiltonian1}) and the typical dispersion relation $E_k = E_0 + (\Delta/2) \cos k$ for the corresponding eigenvalues [\onlinecite{bloch}]. Now, in the presence of disorder introduced by blocks of the $\sqcap$ type, we have the modification

\bea
H_{D}= E_0 + \frac{\Delta}{4} \sum_{n=1}^{N-1}\{ \left[1+\epsilon\sigma_p(n)\right]|n\>\<n+1|  + \rm{h.c.} \},
\label{hamiltonian2}
\eea
where $\sigma_p(n)$ takes the values $0$ and $1$ randomly as a function of the site $n$. The parameter $p$ denotes the ratio of $\sqcap$ to $\sguide$ blocks with $0<p<1/2$. We further impose the constraint that if $\sigma_p(n)=1$ then $\sigma_p(n+1)=0$ in order to avoid self-intersection of the array.

The resulting dispersion relation is depicted in Fig.~\ref{random} for parameters $E_t=1$, $\Delta=4$ and $\epsilon=0.1$. This set of parameters represents a strong perturbation in comparison with the numerically obtained relation $(\Delta \epsilon)/(2 E_t) = 0.142-0.125$ given by the difference between the black and the red curve in Fig.~\ref{levels}(b). With this we show that the claimed robustness in a disordered waveguide is actually stronger. Inspection of Fig.~\ref{random} reveals that the bandwidth increases in small amounts with increasing disorder $p$. The states at the edge of the band suffer the greatest modification, whereas the presence of a conical point located at the center of the band is immune to disorder. From the point of view of symmetry, the existence of a conical point is protected by the fact that all corners are equal, although they are not equally connected. Therefore, off-diagonal perturbations do not modify the {\it swapping\ }symmetry of corners. 

\subsubsection{Perturbative approach}

In order to describe the widening of the bands and the edge effects analytically, we find the corrections to the spectrum by using first order perturbation theory in $\epsilon$. To this end, we first approximate our unperturbed solutions by Bloch waves (finite size effects shall be treated as $O(1/N)$ corrections). We have unperturbed eigenvectors

\bea
| \tilde m \> = \frac{1}{\sqrt{N}}\sum_{n=1}^{N} \exp \left(-i n \frac{2\pi \tilde m}{N} \right) |n\> + O(1/N),
\label{bloch}
\eea
where $\tilde m$ is an integer. The Bloch quasi momentum is recovered in the limit $N \rightarrow \infty$ in the form $2\pi \tilde m / N \rightarrow k$. The
energy shifts are given by
\bea
\Delta^{(1)} E_{\tilde m} & = &\epsilon \<  \tilde m  |  V  | \tilde m  \>  \nonumber \\
&= & \epsilon \frac{\Delta}{4}  \<  \tilde m  | \sum_{n=1}^{N-1}\{ \sigma_p(n) |n\>\<n+1|  + \mbox{h.c.} \}  | \tilde m  \>  \nonumber \\
&= &  \epsilon \frac{\Delta}{2} \cos \left( \frac{2 \pi \tilde m}{N} \right) \sum_{n=1}^{N-1} \{ \sigma_p(n) +  \sigma_p(n+1) \} \nonumber \\
&& + O(1/N^{3/2}) \nonumber \\
&= &\frac{ \epsilon \Delta N_p}{N}  \cos \left( \frac{2 \pi \tilde m}{N} \right) + O(1/N^{3/2}),
\label{shift}
\eea
where $N_p$ is the number of perturbed sites with $\sqcap$ couplings, computed by summing over all $\sigma_p(n)$. For a large number of realizations, the ratio $N_p/N$ tends to $p$ and we may write

\bea
\Delta^{(1)} E_{\tilde m} \approx p \epsilon \Delta  \cos \left( \frac{2 \pi \tilde m}{N} \right).
\label{shiftp}
\eea
Interestingly, at the center of the band $\tilde m= N /4$ and the corrections vanish to leading order independently of $p$. Therefore, the conical points are protected. At any other region of the band, the shifts increase linearly with the disorder parameter $p$. The corrections become significant when (\ref{shiftp}) has a maximum, \ie at the edge of the band corresponding to $\tilde m=0$. The largest correction possible is therefore $p \epsilon \Delta $. This formula for the largest shift has an upper bound of $1.7 \times 10^{-2} E_t$ when $p=0.5$ and with a maximum value of the perturbation $\Delta \epsilon / 2 = 0.017 E_t$ at separation distance $L=d/2$.  

In order to discuss the predictive power of (\ref{shiftp}), we can estimate the numerical energy shift of the blue peaks in Fig.~\ref{G1}(b) with respect to the dashed curve. For a separation parameter $L=2.5d$, the difference of level splittings is of the order of $10^{-3}E_t$, as can be seen from the data plotted in Fig.~\ref{levels}(b). For maximal disorder we have again $p=0.5$, leading to corrections at the edge of the band  $\Delta^{(1)} E_t = p \epsilon \Delta \sim 10^{-3}E_t$, in accordance with the numerical value.

\subsubsection{Wavefunctions and off-diagonal disorder}

When it comes to the discussion of localization, we should not ignore the fact that off-diagonal disorder in tight-binding arrays has been addressed before [\onlinecite{economou}], where a vanishing value of the transmission has been reported. The statistical transmission coefficient computed there, results in a decreasing function of the size of the system multiplied by the strength of some random potential. Although our systems admit a wide range of configurations, our random potential in (\ref{hamiltonian2}) is {\it perturbative.\ }In the light of this result, localization lengths defined in connection with transmission coefficients are inversely proportional to our parameters $\epsilon$ and $p$. As we have proven that $\Delta \epsilon / 2 < 0.017 E_t$ for all separation distances $L$, we can be sure that chains with thousands of corners do not feel the localization effects aforementioned. Moreover, we have used a dimensionless conductance in terms of the $S$ matrix. Such quantity provides a more realistic approach to the transport properties, as it is directly connected to the numerical solutions of the scattering problem in a wire and it gives useful information for each realization of the system.

%

\section{Summary and outlook}

Let us summarize our results. We have studied numerically two models of waveguides with corners and segments with random orientations. Immunity to such disorder is a counterintuitive effect in the context of waves, and we verified its validity by showing that the lowest conductance band of the system is similar to that of a one-dimensional crystal. A theoretical description of the effects in question was given in terms of trapped modes in bent waveguides and their coupling in a tight-binding scheme. Finally, we showed that severely disordered configurations can be treated perturbatively at low energies. This enabled us to estimate the small effects at the edges of the conductance bands, while the immunity of conical (or Dirac) points was confirmed at the center of such bands. 

Our study can be extended to other systems following similar principles. For example, upgrading to two-dimensional systems seems possible, as it only requires the existence of trapped states in cross-wires [\onlinecite{schult}]. We propose the emulation of atomic centers in a monolayer obtained by replicating the system periodically. In connection with particle statistics, we should mention that a canonical second quantization scheme can be proposed on our disordered one-dimensional lattice by promoting the localized or atomic states to field operators. Both fermionic and bosonic schemes are possible. This opens the possibility of describing the propagation of independent electrons in a quantum wire, charge carriers in a medium with an effective wave equation or photons of a fixed polarization and very low frequency.

\section*{Acknowledgments}

The authors are indebted to T. H. Seligman for useful discussions. E. S. is grateful to PROMEP for financial support under project $103.5/12/4367$.


\begin{thebibliography}{99}



\bibitem{veselago}
V. Veselago, \etal
Journal of Computational and Theoretical Nanoscience, {\bf 3}, 1 (2006).

\bibitem{7}
E. Yablonovitch, 
Phys. Rev. Lett. {\bf 58}, 2059 (1987).

\bibitem{10}
J. P. Albert, \etal
Optical and Quantum Electronics, {\bf 34}, 251 (2002).

\bibitem{acoustic}
S. Guenneau, \etal
New J. Phys. {\bf 9}, 399 (2007).

\bibitem{sadurni2} 
E Sadurn\'i, T H Seligman, and F Mortessagne,
New. J. Phys. {\bf 12}, 053014 (2010).

\bibitem{9}
A. J. Heeger, \etal
Rev. Mod. Phys. {\bf 60}, 781 (1988).

\bibitem{inguscio}
F. S. Cataliotti, L. Fallani, F. Ferlaino, C. Fort, P. Maddaloni, and M. Inguscio,
J. Opt. B {\bf 5}, S17 (2003).

\bibitem{open}
B. Dietz, T. Friedrich, M. Miski-Oglu, A. Richter, T. H. Seligman, and K. Zapfe,
Phys. Rev. E {\bf 74}, 056207 (2006);
B. Dietz, T. Friedrich, M. Miski-Oglu, A. Richter, F. Schafer, and T. H. Seligman,
Phys. Rev. E {\bf 80}, 036212 (2009).

\bibitem{mello} 
P. A. Mello and N. Kumar, 
{\it Quantum Transport in Mesoscopic Systems} (Oxford University Press, Oxford, 2004).

\bibitem{stoeckmann}
H. J. St\"ockmann,
{\it Quantum Chaos: An Introduction} (Cambridge University Press, Cambridge, 1999).

\bibitem{mehta}
M. L. Mehta,
{\it Random Matrices} (Elsevier, Oxford, 2004).

\bibitem{anderson}
E. Abrahams, ed., 
{\it 50 years of Anderson localization} (World Scientific, London, 2010).

\bibitem{resonators1}
G. Huillard, \etal 
Phys. Rev. E {\bf 84}, 016602 (2011).

\bibitem{resonators2} 
D. Laurent, O. Legrand, P. Sebbah, C. Vanneste, and F. Mortessagne,
Phys. Rev. Lett. {\bf 99}, 253902 (2007).

\bibitem{schult}
R. Schult, D. G. Ravenhall, and H. W. Wyld,
Phys. Rev. B {\bf 39}, 5476 (1989).

\bibitem{sadurni}
E. Sadurn\'i and W. P. Schleich,
AIP Conf. Proc. {\bf 1323}, 283 (2010).

\bibitem{elasticity}
L. D. Landau and E. M. Lifshitz, 
{\it Theory of Elasticity} (Pergamon Press, Oxford, 1986).

\bibitem{landauer}
R. Landauer, 
Phil. Mag. {\bf 21}, 863 (1970). 

\bibitem{imri}
Y. Imry and R. Landauer, 
Rev. Mod. Phys. {\bf 71}, S306 (1999).

\bibitem{bu}
M. B\"uttiker, Y. Imry, R. Landauer, and S. Pinhas, 
Phys. Rev. B {\bf 31}, 6207 (1985). 

\bibitem{conformal} 
R. Schinzinger and P. Laura, 
{\it Conformal Mapping: Methods and Applications} (Dover, New York, 2003).

\bibitem{gradshteyn} 
I. Gradshteyn and I. Ryzhyk, 
{\it Tables of integrals, series and Products}, seventh edition (Academic Press, Amsterdam, 2007).

\bibitem{exnerl} 
P. Exner, P. Seba, and P. Stovicek, 
Czech. J. Phys. B {\bf 39}, 1181 (1989).

\bibitem{bloch} 
F. Bloch, 
Z. Phys. {\bf 52} 555 (1928).

\bibitem{economou} 
C. M. Soukoulis and E. N. Economou
Phys. Rev. B {\bf 24}, 10 (1981).


\end{thebibliography}
\end{document}